\documentclass[aps,prB,preprint,superscriptaddress]{revtex4-1}
\usepackage{amsmath,amssymb}
\usepackage{graphicx}
\usepackage{color}
\usepackage{ulem}
\draft

\begin{document}
\makeatletter
\@ifundefined{acknowledgements}{\newenvironment{acknowledgements}{\begin{acknowledgments}}{\end{acknowledgments}}}{}
\makeatother

\title{Strong intervalley mixing between copropagating quantum Hall edge channels in a silicon MOSFET}
\author{Gento Yamahata}
\email{E-mail: gento.yamahata@ntt.com}
\affiliation{Basic Research Laboratories, NTT, Inc., 3-1 Morinosato Wakamiya, Atsugi-shi, Kanagawa, 243-0198, Japan }

\author{Takase Shimizu}
\affiliation{Basic Research Laboratories, NTT, Inc., 3-1 Morinosato Wakamiya, Atsugi-shi, Kanagawa, 243-0198, Japan }

\begin{abstract}
Copropagating quantum Hall edge channels provide a promising platform for compact electron interferometry and flying quantum states. In silicon, the valley degree of freedom offers a natural alternative to spin-resolved edge channels because spin-flip scattering is strongly suppressed by the weak spin-orbit interaction. Here, we investigate interchannel transitions between copropagating valley edge channels in a double-layer-gated silicon metal-oxide-semiconductor field-effect transistor. With the bulk filling factor set to \(\nu=2\), two spin-polarized valley edge channels are brought into close proximity near a depleted side gate. We observe strong intervalley mixing, with a transition probability close to \(1/2\), indicating nearly complete equilibration between the two valley edge channels. In contrast, interchannel transport between edge channels with different spin orientations shows negligible transition probability, consistent with suppressed spin-flip scattering in silicon. These results demonstrate that intervalley coupling at a Si/SiO\(_2\) interface can provide a beam-splitter-like operation for copropagating valley edge channels, establishing a key building block toward compact silicon quantum Hall interferometers.
\end{abstract}

\maketitle

\section{Introduction}
Silicon metal-oxide-semiconductor field-effect transistors (MOSFETs) have played a central role in pioneering studies of electron transport in two-dimensional electron systems, culminating in the discovery of the integer quantum Hall effect. \cite{Ando1982,Klitzing1980} Subsequent experimental work on quantum Hall edge transport developed mainly in GaAs/AlGaAs heterostructures, where high mobility and electrostatic gate control enabled detailed studies of edge-channel transport, selective population, and equilibration. \cite{Haug1988,Komiyama1989,Muller1990,Muller1992} As a result, quantum Hall edge transport in silicon has remained much less explored than in GaAs, despite the presence of an additional valley degree of freedom unique to silicon.

Quantum Hall edge channels have also been explored as a platform for flying quantum states. Electronic Mach--Zehnder interferometers based on quantum Hall edge channels were first demonstrated in GaAs/AlGaAs heterostructures, establishing electron interferometry with chiral edge states.\cite{Ji2003} To overcome the limited scalability of conventional architectures, an alternative interferometric scheme based on copropagating edge channels was proposed.\cite{Giovannetti2008} Such a concept was experimentally demonstrated using spin-resolved edge channels under current imbalance,\cite{Deviatov2011} and using magnetic nanofinger beam splitters,\cite{Karmakar2011,Karmakar2015} and has recently been implemented in a more compact architecture using beam splitters based on spin-flip tunneling at sharp gate-defined corners mediated by the local spin-orbit interaction.\cite{Shimizu2020,Shimizu2023PRApp,Shimizu2025MZI,Iyoda2025PRB} The same approach is not naturally transferable to silicon, where spin-orbit coupling is weak and spin-flip scattering between spin-resolved edge channels is expected to be strongly suppressed. A promising alternative is to employ valley edge channels as the interferometric paths. In such an architecture, sufficiently strong intervalley mixing could provide an efficient beam-splitter mechanism.

Previous studies have investigated spin- and valley-resolved edge transport in silicon quantum Hall systems. An early study on Si-MOSFET quantum Hall edge channels demonstrated that spin-flip scattering is strongly suppressed, reflecting the weak spin-orbit interaction in silicon.\cite{Son1992} Subsequent studies of high-mobility Si/SiGe quantum Hall systems demonstrated that spin-resolved edge-channel transport is governed by the weak spin-orbit interaction in silicon, resulting in strongly suppressed spin-flip scattering.\cite{Hamaya2006,Hamaya2007} Valley-resolved edge transport has also been investigated in Si/SiGe, where intervalley equilibration remains incomplete over micrometer-scale transport distances.\cite{Sugihara2008} However, the corresponding intervalley transport in Si-MOSFETs has not yet been clarified. Because the interface properties and electrostatic environment in Si-MOSFETs differ from those in Si/SiGe heterostructures, intervalley coupling in Si-MOSFETs cannot be inferred directly from previous Si/SiGe studies.

Valley physics has also become a central topic in silicon spin-qubit devices, where large and electrically tunable valley splitting is essential for high-fidelity qubit operation.\cite{Buterakos2021,Burkard2023} To better understand valley physics, silicon quantum Hall systems have become an important platform for quantitatively investigating valley splitting and its underlying physics.\cite{Wuetz2020,Lodari2022,Aubergier2025} Investigating valley transport in quantum Hall edge channels contributes to a deeper understanding of valley physics in silicon.

In this work, we investigate intervalley transport between copropagating valley edge channels in a double-layer-gated Si MOSFET. By using local gates, we form a geometry in which two spin-polarized valley edge channels approach each other near a depleted side gate. We find that intervalley mixing is sufficiently strong to produce nearly complete equilibration between the two valley edge channels. In contrast, transitions between edge channels belonging to different spin branches remain negligible. These results demonstrate that Si-MOSFETs can provide the strong intervalley mixing required for valley beam splitters and establish an important building block toward compact valley-based electron interferometers.

\section{Device and measurement setup}
\begin{figure}
\begin{center}
\includegraphics[pagebox=artbox]{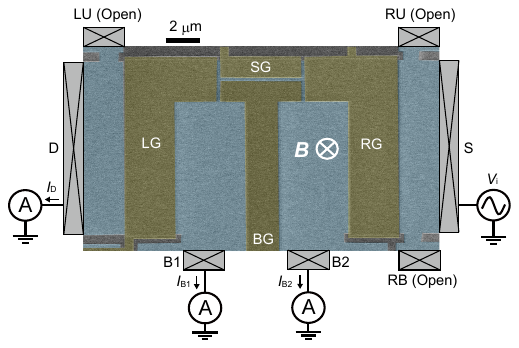}
\end{center}
\caption{Device and measurement setup. A scanning electron micrograph of the double-layer-gated Si MOSFET is overlaid with the measurement circuit used for the intervalley transport measurements. The scanning electron micrograph was taken before fabrication of the global upper gate, which covers the entire device region. An ac excitation \(V_{\mathrm{i}}\) was applied to S. The currents \(I_D\), \(I_{B1}\), and \(I_{B2}\) were converted to voltages by NF CA5351 programmable current amplifiers, detected using an NF LI5650 lock-in amplifier, and the lock-in output voltages were recorded with a Keysight 3458A digital multimeter. The magnetic field \(B\) was applied into the plane of the figure.}
\label{fig:device}
\end{figure}

Figure~\ref{fig:device} shows the device and measurement configuration. The device is a silicon MOSFET with two layers of polycrystalline silicon gates fabricated on a silicon-on-insulator substrate. The device was fabricated using the same fabrication process as in our previous study on silicon single-electron pumps.\cite{Yamahata2025NanoLett} A micrometer-scale silicon mesa was first defined on a silicon-on-insulator substrate by electron-beam lithography and dry etching. A 30-nm-thick thermal oxide was then grown, followed by an \(n\)-type heavily doped lower polycrystalline silicon gate layer patterned by electron-beam lithography and dry etching. After deposition of a 50-nm-thick interlayer silicon oxide layer by chemical vapor deposition, an upper polycrystalline silicon gate covering the entire device region was formed by photolithography and dry etching. Phosphorus-doped contact regions were formed by ion implantation using the upper gate as a mask, and finally aluminum ohmic electrodes were formed on the contact regions.

The silicon two-dimensional electron region is connected to four electrical terminals, S, D, B1, and B2, which can be used for voltage excitation and current measurement. The terminals LU, RU, and RB were left open throughout this work. Figure~\ref{fig:device} illustrates the measurement configuration used for the intervalley transport measurements, in which an ac voltage was applied to terminal S and the resulting currents were measured at terminals D, B1, and B2. The lower gates LG, RG, SG, and BG locally control the filling factors underneath them, whereas the global upper gate UG controls the bulk filling factor. Specifically, LG and RG tune the edge-channel populations at the entrance and exit, BG is adjusted to match the bulk filling factor, and SG is either tuned to the bulk filling factor or biased negatively to deplete the region underneath it.

Measurements were performed in a dilution refrigerator at a temperature of 280 mK. The magnetic field was fixed at \(B=12.92\) T. We applied an ac excitation of \(V_{\mathrm{i}}=27.25\,\mu\mathrm{V}_{\mathrm{rms}}\) at 33 Hz. The measurements were carried out using standard low-frequency lock-in techniques.

\section{Gate control of valley edge channels}
\begin{figure}
\begin{center}
\includegraphics[pagebox=artbox]{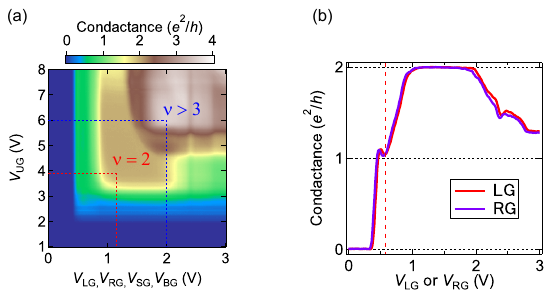}
\end{center}
\caption{Gate control of the quantum Hall filling factor. (a) Two-terminal conductance measured between S and D as a function of the upper-gate voltage \(V_{\mathrm{UG}}\) and the common lower-gate voltage applied to LG, RG, SG, and BG. The red dashed lines mark the gate-voltage setting used for the intervalley transport experiment. The blue dashed lines mark the setting used for the higher-filling-factor control experiment. (b) Two-terminal conductance measured between S and D as a function of \(V_{\mathrm{LG}}\) or \(V_{\mathrm{RG}}\), with \(V_{\mathrm{UG}}=3.9\) V and the remaining lower-gate voltages fixed at 1.15 V, corresponding to a bulk filling factor of \(\nu=2\). The dashed vertical line indicates the voltage used to tune the regions under LG and RG close to \(\nu=1\).}
\label{fig:plateau}
\end{figure}

We first characterize the filling factors controlled by the upper and lower gates. Figure~\ref{fig:plateau}(a) shows the two-terminal conductance measured between S and D, while terminals B1 and B2 were left open. The vertical axis is the upper-gate voltage \(V_{\mathrm{UG}}\), and the horizontal axis is the common lower-gate voltage applied simultaneously to LG, RG, SG, and BG. The conductance is corrected for parasitic series resistance as described in Appendix~\ref{app:series}. A well-defined plateau corresponding to \(\nu=2\) is observed. At \(V_{\mathrm{UG}}=3.9\) V and \(V_{\mathrm{LG}}=V_{\mathrm{RG}}=V_{\mathrm{SG}}=V_{\mathrm{BG}}=1.15\) V, all regions are set to \(\nu=2\), as indicated by the red dashed lines. We attribute this state to two spin-polarized edge channels with different valley indices. At \(V_{\mathrm{UG}}=6\) V and lower-gate voltages of 2 V, the filling factor exceeds 3, as indicated by the blue dashed lines. In this regime, edge channels with spin orientations different from those at \(\nu=2\) contribute to transport.

Figure~\ref{fig:plateau}(b) shows the two-terminal conductance measured between S and D as a function of \(V_{\mathrm{LG}}\) or \(V_{\mathrm{RG}}\) with the bulk fixed at \(\nu=2\). A clear \(2e^2/h\) plateau is obtained for \(\nu=2\), and a conductance near \(e^2/h\) is observed around the \(\nu=1\) setting, although it is less ideal. We therefore use \(V_{\mathrm{LG}}=V_{\mathrm{RG}}=0.58\) V, marked by the vertical dashed line, to tune the regions under LG and RG close to \(\nu=1\) while keeping the bulk at \(\nu=2\). This configuration enables intervalley transport between the two spin-polarized valley edge channels to be investigated.

\section{Interchannel transitions between copropagating quantum Hall edge channels}
\begin{figure*}
\begin{center}
\includegraphics[pagebox=artbox]{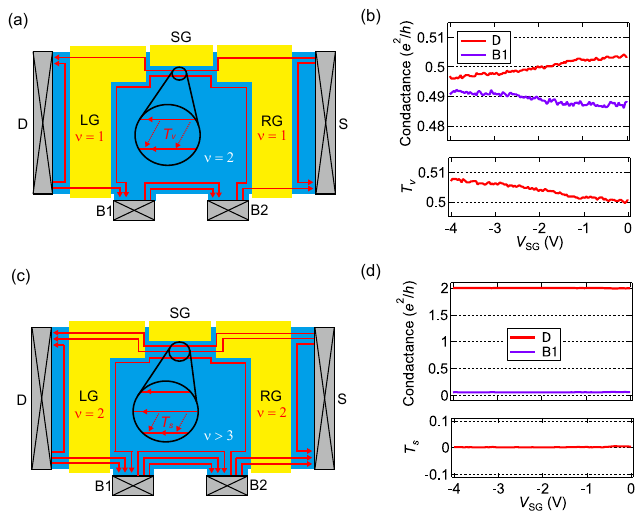}
\end{center}
\caption{Transition measurements between copropagating quantum Hall edge channels. (a) Schematic of the edge-channel configuration for the intervalley transport measurement. The regions under LG and RG are tuned near \(\nu=1\), while the bulk is at \(\nu=2\). Near the depleted SG region, the two spin-polarized valley edge channels approach each other, and the transition probability is denoted by \(T_v\). (b) Upper panel: conductances measured at D and B1 as a function of \(V_{\mathrm{SG}}\). Lower panel: intervalley transition probability \(T_v\) extracted from the measured currents. (c) Schematic of the edge-channel configuration with different spin orientations. The regions under LG and RG are set to \(\nu=2\), while the bulk is at \(\nu>3\). (d) Conductances and extracted transition probability \(T_s\).}
\label{fig:transition}
\end{figure*}

We now investigate intervalley transitions between the two valley edge channels. The measurement configuration is the same as that shown in Fig.~\ref{fig:device}, and the edge-channel configuration is illustrated schematically in Fig.~\ref{fig:transition}(a). The regions under LG and RG are set near \(\nu=1\), and the bulk is set to \(\nu=2\). A negative voltage is applied to SG, depleting the region underneath the gate. In this geometry, the outer edge channel injected from S and the inner edge channel connected to B2 run close to each other near SG. We define \(T_{\mathrm{v}}\) as the transition probability from the outer valley edge channel to the inner valley edge channel in this region, as indicated in the inset of Fig.~\ref{fig:transition}(a).

In this configuration, the current measured at B2 is nearly zero. This indicates that the edge channel running along RG from B2 toward SG is equilibrated and does not inject a measurable current into the active edge path. Figure~\ref{fig:transition}(b) shows the conductances measured at D and B1 as a function of \(V_{\mathrm{SG}}\), together with the transition probability extracted using the procedure described in Appendix~\ref{app:transition}. The conductances at D and B1 are both close to \(0.5e^2/h\), and the extracted transition probability is close to \(T_{\mathrm{v}}=0.5\). These results demonstrate strong intervalley mixing between the two valley edge channels in the Si-MOSFET. Although a small and reproducible dependence on \(V_{\mathrm{SG}}\) is observed, the transition probability remains close to 0.5 over the entire measured range. The origin of this weak gate dependence is presently unclear and may be related to the nonideal \(\nu=1\) plateau. A more quantitative investigation using devices exhibiting a better-developed \(\nu=1\) plateau will be the subject of future work. Nevertheless, the present results clearly demonstrate sufficiently strong intervalley mixing for the realization of a valley beam splitter.

We next investigate interchannel transport between edge channels with different spin orientations. The measurement configuration is the same as that shown in Fig.~\ref{fig:device}, and the edge-channel configuration is illustrated schematically in Fig.~\ref{fig:transition}(c). The regions under LG and RG are set to \(\nu=2\), whereas the bulk is set to \(\nu>3\). In this case, the additional inner edge channel in the bulk has a spin orientation different from that of the outer \(\nu=2\) edge channels, and an interchannel transition therefore requires a spin flip.

Figure~\ref{fig:transition}(d) shows the conductances measured at D and B1, together with the transition probability extracted using the procedure described in Appendix~\ref{app:transition}. As in the intervalley transport measurement, the current measured at B2 is nearly zero. In contrast to Fig.~\ref{fig:transition}(b), however, the conductance at B1 is also nearly zero, while the conductance at D remains close to \(2e^2/h\). The extracted transition probability \(T_{\mathrm{s}}\), as indicated in the inset of Fig.~\ref{fig:transition}(c), is nearly zero over the measured range of \(V_{\mathrm{SG}}\). This indicates that spin-flip inter-edge-channel scattering is strongly suppressed in the present Si-MOSFET. The suppression of \(T_{\mathrm{s}}\) is consistent with earlier studies of silicon quantum Hall systems, in which spin-flip inter-edge-channel scattering was found to be strongly suppressed due to the weak spin-orbit interaction in silicon. \cite{Son1992,Hamaya2006,Hamaya2007}
 
\section{Discussion}
\begin{figure}
\begin{center}
\includegraphics[pagebox=artbox]{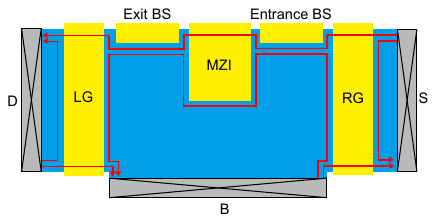}
\end{center}
\caption{Conceptual valley Mach--Zehnder interferometer using copropagating valley edge channels. Two intervalley-mixing regions act as the entrance and exit beam splitters (BSs). The present observation of strong intervalley mixing establishes the key building block for such a device.}
\label{fig:mzi}
\end{figure}

The strong intervalley transition observed here suggests a route to a valley beam splitter in silicon. Figure~\ref{fig:mzi} illustrates a possible Mach--Zehnder interferometer based on copropagating valley edge channels. Two gate-controlled intervalley-mixing regions would serve as the entrance and exit beam splitters. Between them, the two valley edge channels would propagate along spatially separated paths, accumulating a relative phase that could be tuned by electrostatic gates or magnetic flux.

Compared with GaAs copropagating spin interferometers, a silicon valley interferometer may not require deliberately sharp bends for spin-orbit-induced spin flips. \cite{Shimizu2023PRApp} Instead, intervalley coupling may arise from the Si/SiO\(_2\) interface when two valley edge channels are brought into close proximity by local electrostatic gates, potentially enabling more compact and simpler device architectures. Furthermore, silicon may exhibit physical phenomena distinct from those in GaAs because the relevant internal degree of freedom, hyperfine environment, and spin-orbit interaction are fundamentally different.

The intervalley mixing observed in the present Si-MOSFET is stronger than that reported in previous Si/SiGe experiments. In Si/SiGe valley-resolved edge transport, intervalley equilibration was observed over micrometer-scale distances, but complete equilibration was not reached within the measured transport length.\cite{Sugihara2008} In contrast, the present Si-MOSFET shows nearly complete equilibration over a comparable length scale. This difference suggests that intervalley coupling is enhanced in the Si-MOSFET structure.

The microscopic origin of the enhanced intervalley coupling remains to be clarified. Possible mechanisms include interface roughness, local disorder, and defect-induced scattering at the Si/SiO\(_2\) interface. Importantly, such scattering potentials are static after device fabrication. Therefore, once a device geometry is defined, the intervalley scattering process is expected to be reproducible for each electron passing through the mixing region. If phase coherence is preserved, this static intervalley coupling could function as a coherent beam splitter for valley edge channels.

Several improvements are still required before a coherent valley interferometer can be realized. In the present device, the \(\nu=1\) quantum Hall state is not as well developed as the \(\nu=2\) state, which limits the ideality of valley-channel initialization and readout. Previous studies have demonstrated that a well-developed \(\nu=1\) quantum Hall state can be achieved in Si-MOSFET structures through optimized device design and operation at higher magnetic fields.\cite{Takashina2007} Further optimization of the device structure and measurement conditions is therefore expected to improve the initialization and readout fidelity of valley edge channels. It will also be necessary to form two spatially separated mixing regions and to characterize phase coherence of the valley superposition during propagation. The present experiment establishes the first of these ingredients: a strong intervalley channel-mixing operation in a Si-MOSFET quantum Hall device.

\section{Conclusion}
We have studied interchannel transitions between copropagating quantum Hall edge channels in a double-layer-gated Si MOSFET. With the bulk set to \(\nu=2\) and the entrance and exit gates tuned near \(\nu=1\), the two spin-polarized valley edge channels are strongly mixed near a depleted side gate, giving a transition probability close to \(1/2\). By contrast, interchannel transport between edge channels with different spin orientations shows negligible transition probability, confirming the strong suppression of spin-flip inter-edge-channel scattering in silicon. These findings show that intervalley coupling at a Si/SiO\(_2\) interface can provide an elementary beam-splitter-like operation for copropagating valley edge channels. They open a path toward compact silicon quantum Hall interferometers and flying valley qubits based on valley edge channels.

\begin{acknowledgements}
The authors thank K. Nishiguchi for valuable discussions and assistance with device fabrication.
\end{acknowledgements}

\appendix
\section{Parasitic-resistance correction}
\label{app:series}

\begin{figure}
\begin{center}
\includegraphics[pagebox=artbox]{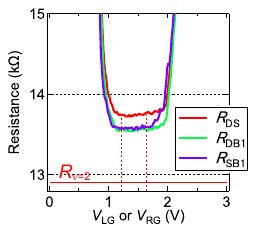}
\end{center}
\caption{Two-terminal resistance used to estimate parasitic series resistances. The resistance was obtained by dividing the excitation voltage \(V_\mathrm{i}\) by the current measured in the three two-terminal configurations: D--S, D--B1, and S--B1. The D--S and D--B1 data are plotted as a function of \(V_{\mathrm{LG}}\), whereas the S--B1 data are plotted as a function of \(V_{\mathrm{RG}}\). The averages were taken over the gate-voltage ranges indicated by the red dashed lines, corresponding to the \(\nu=2\) quantum Hall plateau in both the bulk and edge regions. The red solid line indicates the quantized resistance \(R_{\nu=2}=h/2e^2\).}
\label{fig:series}
\end{figure}
We corrected the conductance data for parasitic series resistances in the device and measurement system. To estimate the relevant resistances, three two-terminal measurements were performed under the condition that both the bulk and edge regions correspond to the \(\nu=2\) quantum Hall plateau. Voltage was applied to D and the current was measured at S, voltage was applied to D and the current was measured at B1, and voltage was applied to S and the current was measured at B1. The measured currents flowing into the electrodes are denoted by \(I_{DS}\), \(I_{DB1}\), and \(I_{SB1}\), respectively. The excitation voltage after the 80-dB attenuator was directly measured using the lock-in amplifier, yielding \(V_{\mathrm{i}}=27.25\,\mu\mathrm{V}_{\mathrm{rms}}\). The corresponding two-terminal resistances, \(R_{DS}=V_{\mathrm{i}}/I_{DS}\), \(R_{DB1}=V_{\mathrm{i}}/I_{DB1}\), and \(R_{SB1}=V_{\mathrm{i}}/I_{SB1}\), are plotted as a function of gate voltage in Fig.~\ref{fig:series}. In the following analysis, the average values over the gate-voltage ranges indicated by the red dashed lines were used. The slightly larger value of \(R_{DS}\) is attributed to the difference between the symmetric S--D geometry and the geometry involving B1.

Let \(R_D\), \(R_S\), and \(R_{B1}\) be the parasitic resistances associated with the corresponding terminals, and let \(R_m\) be the parasitic resistance of the measurement system. The electric potentials at the terminals are denoted by \(V_D\), \(V_S\), and \(V_{B1}\). For the \(\nu=2\) plateau, the Landauer--B{\"u}ttiker equations \cite{Buttiker1988} for the three configurations are
\begin{align}
I_{DS}
&=\frac{2e^2}{h}\left(V_S-V_D\right) \\
&=\frac{2e^2}{h}\left[V_{\mathrm{i}}-(R_D+R_S+R_m)I_{DS}\right],
\label{eq:ids}
\\[3pt]
I_{DB1}
&=\frac{2e^2}{h}\left(V_D-V_{B1}\right) \\
&=\frac{2e^2}{h}\left[V_{\mathrm{i}}-(R_D+R_{B1}+R_m)I_{DB1}\right],
\label{eq:idb1}
\\[3pt]
I_{SB1}
&=\frac{2e^2}{h}\left(V_S-V_{B1}\right) \\
&=\frac{2e^2}{h}\left[V_{\mathrm{i}}-(R_S+R_{B1}+R_m)I_{SB1}\right].
\label{eq:isb1}
\end{align}
Equation~\eqref{eq:ids} gives
\begin{equation}
R_D+R_S+R_m=R_{DS}-\frac{h}{2e^2}=856\,\Omega .
\label{eq:rds}
\end{equation}
The two-terminal conductance in Fig.~\ref{fig:plateau} was therefore corrected using
\begin{equation}
G_{DS}=\left[\frac{V_{\mathrm{i}}}{I_{DS}}-(R_D+R_S+R_m)\right]^{-1}.
\label{eq:gds}
\end{equation}
The conductance measured at D in Fig.~\ref{fig:transition} was corrected in the same way.

From Eq.~\eqref{eq:isb1}, the total parasitic resistance associated with the B1 measurement is obtained as
\begin{equation}
R_{B1}+R_S+R_m=R_{SB1}-\frac{h}{2e^2}=650\,\Omega .
\label{eq:rb1s}
\end{equation}
Using this value, the corrected conductance is
\begin{equation}
G_{B1S}=\left[\frac{V_{\mathrm{i}}}{I_{B1S}}-(R_{B1}+R_S+R_m)\right]^{-1},
\label{eq:gb1s}
\end{equation}
where \(I_{B1S}\) denotes the current measured at B1 when the excitation is applied to S. The corrected conductance was used for the B1 conductance shown in Fig.~\ref{fig:transition}(b,d).

\section{Extraction of the transition probability}
\label{app:transition}
We define \(T_{\mathrm{v}}\) as the transition probability from the outer valley edge channel to the inner valley edge channel near SG, as illustrated in Fig.~\ref{fig:transition}(a). Let \(I_S\), \(I_D\), \(I_{B1}\), and \(I_{B2}\) denote the currents flowing into the corresponding electrodes. Current conservation gives
\begin{equation}
I_S+I_D+I_{B1}+I_{B2}=0.
\end{equation}
The Landauer--B{\"u}ttiker relation is
\begin{align}
I_D
&=(1-T_{\mathrm{v}})\frac{e^2}{h}\left(V_S-V_D\right) \nonumber\\
&=(1-T_{\mathrm{v}})\frac{e^2}{h}\left \{ V_{\mathrm{i}}+R_SI_S-(R_D+R_m)I_D\right \} \nonumber\\
&=(1-T_{\mathrm{v}})\frac{e^2}{h}\mathcal{D},
\label{eq:td}
\end{align}
where 
\begin{equation}
\mathcal{D}=V_{\mathrm{i}}-(R_D+R_S+R_m)I_D
-R_S(I_{B1}+I_{B2}).
\end{equation}
Solving Eq.~\eqref{eq:td} gives
\begin{equation}
T_{\mathrm{v}}
=1-\frac{h}{e^2}\frac{I_D}{\mathcal{D}} .
\label{eq:tvalley}
\end{equation}

Together with Eqs.~\eqref{eq:ids}--\eqref{eq:isb1}, the transition probability \(T_{\mathrm{v}}\) is expressed as a function of the measured currents and \(R_m\). The finite \(R_m\) is attributed mainly to the current amplifier. A separate measurement of the current-amplifier input impedance using a fixed \(10\,\mathrm{k}\Omega\) resistor indicated that \(R_m\) is on the order of \(50\,\Omega\) at 33 Hz. In Fig.~\ref{fig:transition}(b), we therefore used \(R_m=50\,\Omega\). Even if \(R_m=100\,\Omega\) is assumed instead, the extracted transition probabilities change by less than 1\%, leaving the conclusions unchanged.

For the spin-flip configuration in Fig.~\ref{fig:transition}(c), the incident edge corresponds to \(\nu=2\). The same analysis is applied after replacing \(e^2/h\) in Eq.~\eqref{eq:td} by \(2e^2/h\). The data in Fig.~\ref{fig:transition}(d) were obtained with \(R_m=50\,\Omega\).

\end{document}